\documentclass[aps,pre,reprint,superscriptaddress,showpacs,preprintnumbers,amsmath,amssymb]{revtex4-1}
\usepackage{graphicx}
\usepackage{siunitx}
\usepackage{amsmath}
\usepackage{amsfonts}
\usepackage{amssymb}

\usepackage{natbib}
\usepackage{xcolor}
\usepackage{url}
\usepackage{bm}

\DeclareSIUnit\Molar{\textsc{m}}

\usepackage{xcolor}
\usepackage{soul}

\begin{document}
\title{Stress-activated Constraints in Dense Suspension Rheology}
%

\author{Abhinendra Singh}
\email{abhinendra@uchicago.edu}
\affiliation{James Franck Institute, University of Chicago, Chicago, Illinois 60637, USA}
\affiliation{Pritzker School of Molecular Engineering, University of Chicago, Chicago, Illinois 60637, USA}
\author{Grayson L. Jackson} 
\affiliation{James Franck Institute, University of Chicago, Chicago, Illinois 60637, USA}
\author{Michael van der Naald}
\affiliation{James Franck Institute, University of Chicago, Chicago, Illinois 60637, USA}
\affiliation{Department of Physics, The University of Chicago, Chicago, Illinois 60637, USA}
\author{Juan J de Pablo} 
\affiliation{Pritzker School of Molecular Engineering, University of Chicago, Chicago, Illinois 60637, USA}
\affiliation{Pritzker School of Molecular Engineering, University of Chicago, Chicago, IlliMaterials Science Division, Argonne National Laboratory, Lemont, Illinois 60439, USA}
\author{Heinrich M. Jaeger}
\affiliation{James Franck Institute, University of Chicago, Chicago, Illinois 60637, USA}
\affiliation{Department of Physics, The University of Chicago, Chicago, Illinois 60637, USA}




\begin{abstract}
Dispersing small particles in a liquid can produce surprising behaviors when the solids fraction becomes large: rapid shearing drives these systems out of equilibrium and can lead to dramatic increases in viscosity (shear thickening) or even solidification (shear jamming). These phenomena occur above a characteristic onset stress when particles are forced into frictional contact.
Here we show  via simulations how this can be understood within a framework that abstracts details of the forces acting at particle-particle contacts into general stress-activated constraints on relative particle movement. We find that focusing on just two constraints, affecting sliding and rolling at contact, can reproduce the experimentally observed shear thickening behavior quantitatively, despite widely different particle properties, surface chemistries, and suspending fluids. Within this framework parameters such as coefficients of sliding and rolling friction can each be viewed as proxy for one or more forces of different physical or chemical origin, while the parameter magnitudes  indicate the relative importance of the associated constraint. In this way, a new link is established that connects features observable in macroscale rheological measurements to classes of constraints arising from micro- or nano-scale properties.
\end{abstract}


\maketitle

Concentrated or ``dense'' suspensions comprising small particles in a suspending liquid are ubiquitous in nature as well as industrial settings ~\cite{Coussot_1997,van2018concrete,blanco2019conching,Brown_2014,Morris_2020}.
Even for simple liquids with completely Newtonian characteristics, i.e. constant viscosity, such suspensions can exhibit strikingly non-Newtonian behaviors under applied shear, such as yielding, shear thinning, shear thickening, and shear-jamming~\cite{Morris_2020,Mewis_2011,Brown_2014,Denn_2018,guazzelli_2018,Singh_2019}. 
All of these behaviors originate from short-ranged forces between particles that  either are mediated by the hydrodynamics of thin liquid layers or involve direct, frictional and possibly also cohesive contact~\cite{Jamali_2019, Comtet_2017, james2018interparticle}.
The strength of these forces depends not only on the particles' physical properties such as shape, stiffness and surface roughness, but also on their surface chemistry and its role during particle-particle contact or through interaction with the surrounding liquid~\cite{Hsu_2018, James_2019,Bourrianne_2020, Naald_2021, Hsu_2021}. 
Given this large set of potentially contributing factors, establishing a predictive link between microscale properties and macroscale observable flow behaviors has been a longstanding problem.

One recent approach, by Guy \emph{et al.}, to address this issue has been to classify the macroscale rheology not by focusing on the details of specific forces, but rather on the general types of constraints that affect relative particle movement~\cite{Guy_2018}. 
The promise of this approach lies in that the physical or chemical origin of any particular particle-particle interaction may matter far less than its net effect on the ability of neighboring particles to move with respect to one another.
What remained to be shown, however, is which specific types of constraints are necessary for quantitative modeling of dense suspension rheology.

Here we demonstrate that quantitative modeling and prediction is indeed possible.  
We focus on shear thickening, perhaps the most remarkable non-Newtonian behavior of dense suspensions, and introduce a new diagnostic framework to classify  rheological flow curves in terms of whether sliding and rolling constraints on particle motion are present. 
This approach provides  insight into why particular particle-scale properties introduce additional constraints while others do not and unlocks a new connection between bulk rheology and nanoscale particle surface properties (Fig. 1A). 

During shear thickening the shear stress $\sigma$ increases faster than the shear rate $\dot\gamma$, leading to a net increase in viscosity $\eta=\sigma/\dot\gamma$.
For low solids volume fractions $\phi$ this increase is mild and occurs continuously as a function of applied shear, but for larger $\phi$ the viscosity can increase abruptly and dramatically when a critical rate $\dot\gamma_c$ is reached, behavior termed discontinuous shear thickening (DST).
For sufficiently large $\phi$ and shear stress $\sigma$, these suspensions can furthermore transform into a shear-jammed (SJ), solid-like state, which ``melts'' back into a fluid once stress is released~\cite{Peters_2016,Seto_2019}.
%
%
Past work has established a strong foundation to understand the evolution of shear thickening toward DST and SJ, yet focused almost exclusively on stress-activated sliding friction, represented by a single coefficient for sliding friction $\mu_s$.

\begin{figure}[t]
\centering
\includegraphics[width=.95\linewidth]{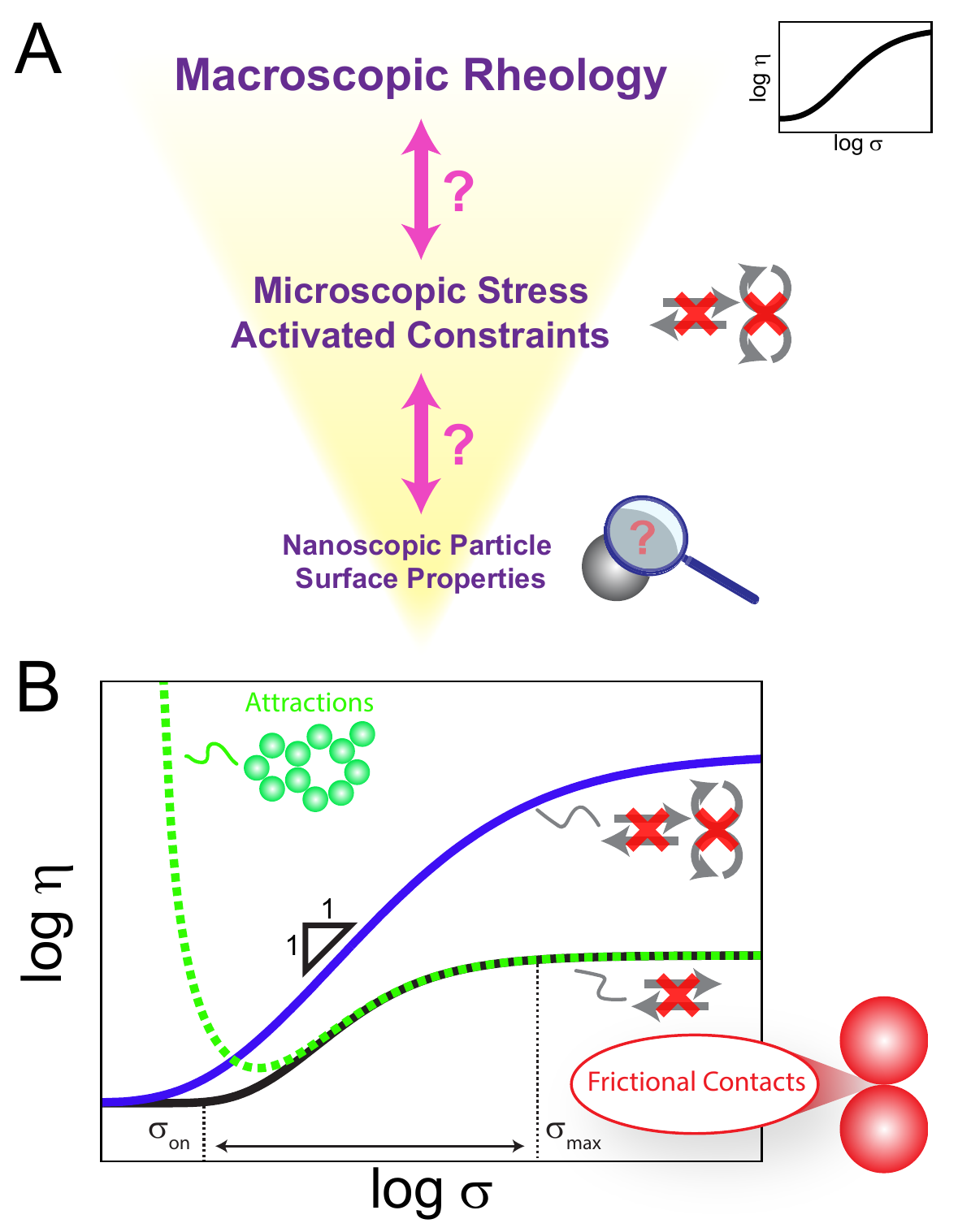}
\caption{(A) Shear thickening is a phenomena that can be understood across a hierarchy of length scales, and unraveling the inter-relationship between macroscopic rheology, microscopic stress-activated constraints that hinder relative particle motion, and nanoscopic particle surface properties poses a major challenge.
Macroscale rheology can become a sensitive probe of nanometer-scale interactions between particle surfaces once the link with microscopic frictional constraints has been established.
(B) How microscopic frictional constraints affect shear thickening. 
Viscosity $\eta$ plotted as function of applied shear stress $\sigma$ at constant volume fraction $\phi$.
Above the critical onset stress $\sigma_{\mathrm{on}}$, ``lubricated'' contacts are starting to become transformed into ``constrained'' frictional ones. At $\sigma_{\mathrm{max}}$ all contacts are constrained and a maximum plateau viscosity is reached.
At this volume fraction, stress-activated sliding constraints alone lead only to continuous shear thickening (\emph{solid black line}).
Attractive central forces typically lead to a yield stress, without affecting shear thickening (\emph{dashed green line}).
Adding stress-activated rolling constraints can lead to discontinuous shear thickening (DST, slope 1) over a wider stress range and with a higher plateau viscosity (\emph{solid blue line}).
}
\label{fig:Sketch}
\end{figure}

In this picture, once the applied shear stress overwhelms the repulsive interparticle potential, unconstrained, hydrodynamically ``lubricated'' contacts transition to ``frictional'' contacts that prevent sliding~\cite{Seto_2013a, Mari_2014, Guy_2015, Ness_2016, Singh_2018, james2018interparticle, Clavaud_2017, Hsu_2018}.
When viscosity $\eta$ is plotted as a function of applied shear stress $\sigma$ (Fig.~\ref{fig:Sketch}B), this increase in the number of frictionally constrained particle-particle contacts manifests as shear thickening that starts at an onset stress $\sigma_{\mathrm{on}}$ and persists up to an upper limit $\sigma_{\mathrm{max}}$, where the system has reached a state with all contacts frictional.
As the solids fraction $\phi$ gets closer to the onset packing fraction for jamming $\phi_J^\mu$, the dependence of $\eta$ on stress  becomes steeper within the shear thickening regime, until DST is reached.
In plots like Fig.~\ref{fig:Sketch}B, DST is identified by a slope $d$(log$\eta$)/$d$(log$\sigma)=1$, i.e., the viscosity  $\eta=\sigma/\dot\gamma_c$ is directly proportional to the stress.
Increasing the sliding friction coefficient $\mu_s$ reduces $\phi_J^\mu$ and thus, for given $\phi$, brings the system closer to jamming.
This in turn steepens the rise in viscosity with stress while at the same time increasing the final viscosity level that is reached in the large stress limit. 

However, if only sliding constraints are considered, the effect of $\mu_s$ on $\phi_J^\mu$ is rather small: it reduces the onset of jamming from $\phi_J^0$ for frictionless ($\mu_s = 0$) particles, which for monodisperse rigid spheres is equal to the random close-packing value $\phi_{\rm{RCP}} \approx 0.64$~\cite{Mari_2014, Wyart_2014,Singh_2018, Guy_2018}, to no lower than $\phi_J^\mu\approx 0.56$  even for $\mu_s = \infty$~\cite{Singh_2018, Mari_2014}.
This poses a serious problem for quantitative prediction, since experiments with rough spherical particles and also with specific chemical surface groups have demonstrated DST for packing fractions so low that the associated $\phi_J^\mu$ lies well below $0.56$ and thus outside the range of current models based on just $\mu_s$~\cite{Laun_1984,james2018interparticle,James_2019,Hsu_2018,hsiao2019experimental,Lootens_2005, Pradeep_2020, Hsu_2021}.
Therefore, additional constraints beyond sliding are needed to properly capture the behavior of real suspensions.
%
%
%

%
%
Taking cues from modeling the rheology of dry granular materials~\cite{Estrada_2011,Ai_2011,Santos_2020,Ding_2007, Dominik_1995}, recent simulations explored how additional stress-activated  constraints on rolling affect shear thickening in dense suspensions~\cite{Singh_2020}.
Interestingly, under the right conditions, already small additions of rolling friction were found to generate significant effects.
At the same volume fraction where suspensions with only sliding constraints exhibit mild shear thickening, adding rolling friction can lead to DST, broaden the stress range over which shear thickening is observed, and increase the viscosity of the frictional state.
In Fig.~\ref{fig:Sketch}B this is shown by comparing the flow curves with (blue) and without (black) rolling friction.
In the limit of infinite sliding and rolling friction the frictional jamming point drops as low as $\phi_J^\mu = 0.37$.
Short-ranged attractive particle-particle interactions that are not stress-activated and give rise to a yield stress can be included by simply adding them~\cite{Singh_2019}, as exemplified by the dashed green trace in Fig.~\ref{fig:Sketch}B for the case without rolling friction.

Taken together, this opens up an opportunity that we explore here: to  model experimental suspension rheology quantitatively and understand how the combination of stress-activated sliding and rolling constraints, expressed in terms of an onset stress for frictional contact and  coefficients for sliding and rolling friction, can be linked to particle-scale properties.

To this end we utilize simulations of rigid spheres that include lubrication, electrostatic repulsion, and frictional sliding and rolling constraints. 
We demonstrate how quantitative detail about the constraints operative over nanoscale distances at particle-particle contacts can be extracted from bulk measurements of $\eta(\sigma)$ as in Fig.~\ref{fig:Sketch}B by comparing the simulations with experimental data for spherical particles.
We first show that a number of commonly studied suspensions that vary widely in their composition, surprisingly have nearly identical frictional constraints and that they primarily affect sliding of contacting particle surfaces.
We term these ``standard'' particle suspensions.
We then use the simulations to predict how the  magnitudes of stress-activated sliding or rolling friction coefficients alter the relative contributions from the associated constraints to the measured $\eta(\sigma)$ curves.
Finally, we illustrate with three examples  how modifications to particle surface roughness or chemistry lead to characteristic deviations from the ``standard'' behavior and how this can be understood in terms of additional stress-activated constraints on particle rolling.

\section*{Results}
%
%
%

\subsection{``Standard'' particle suspensions}

\begin{figure*}[!htbp]
\centering
\includegraphics[width=0.85\textwidth]{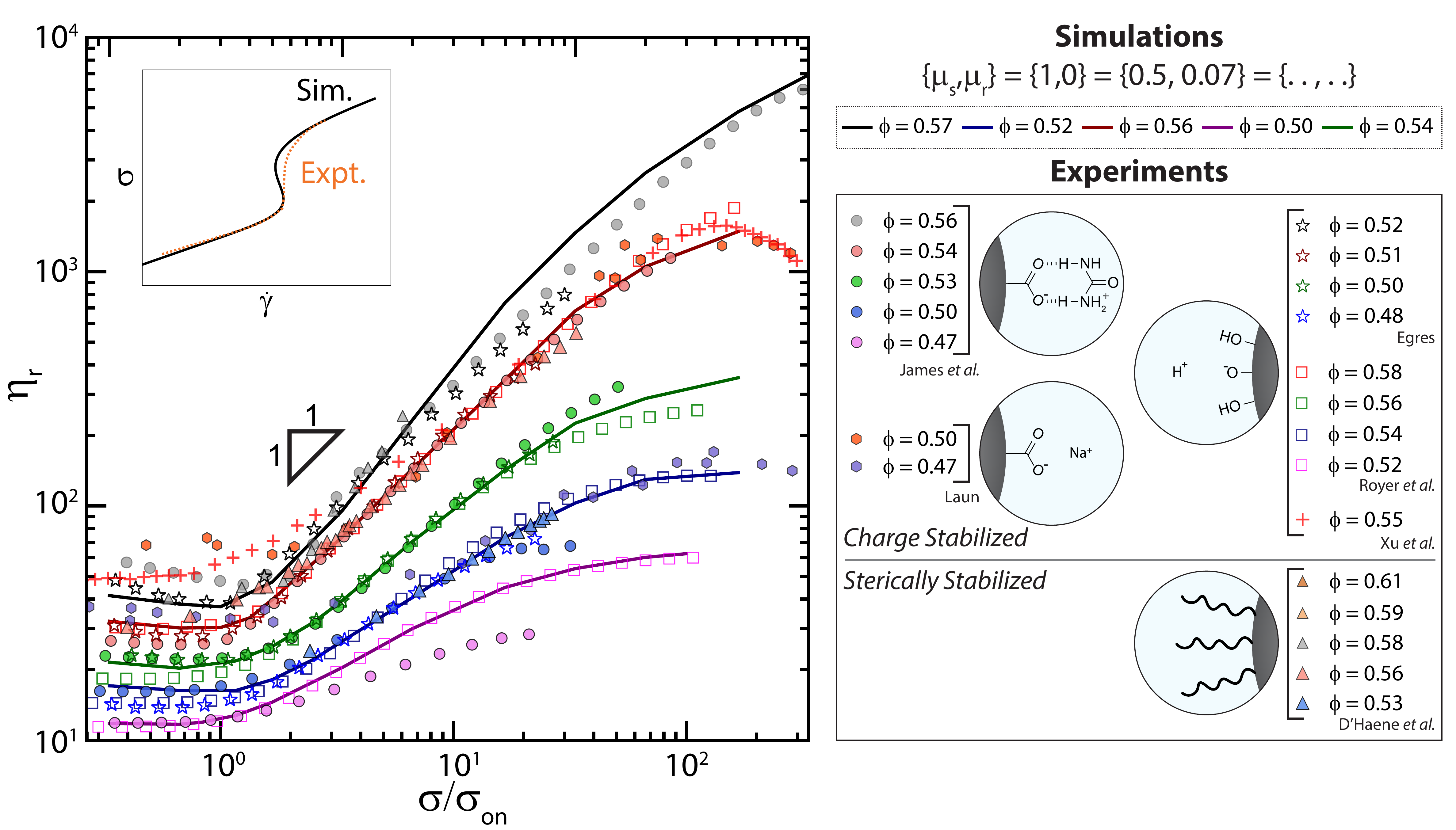}
\caption{
Shear thickening of ``standard'' particle suspensions show nearly identical shear thickening after scaling by the onset stress $\sigma_{\mathrm{on}}$. 
As there are small discrepancies in the experimentally-reported volume fractions, data with similar thickening are grouped by color.
See Table~\ref{tab:Table_1} for details regarding particle size, solvent, and onset stress.
Solid lines denote simulation data for $\{\mu_s,\mu_r\} = \{0.5,0.07\}$ and Debye length $\lambda/a=0.01$ at various volume fractions $\phi$ as mentioned.
(\emph{Inset}) Beyond the onset of DST $\phi_c<\phi<\phi_J^\mu$, simulations (\emph{solid black line}) capture non-monotonic flow curves while experimental measurements cannot (\emph{orange dashed line}).
}
\label{fig:Repulsive_Frictional}
\end{figure*}
In Fig.~\ref{fig:Repulsive_Frictional} we compare the shear thickening behavior across a series of  suspension types and volume fractions:  sterically stabilized poly(methyl methacrylate) (PMMA) particles in dioctyl phthalate ~\cite{DHaene_1993}, charge stabilized silica particles in aqueous glycerol~\cite{Royer_2016} or PEG-200~\cite{Egres_2005a}, glass beads in poly(dimethylsiloxane) (PDMS)~\cite{Xu_2020}, and carboxylate-coated particles in aqueous solutions~\cite{Laun_1984,james2018interparticle}.
While the differences in particle size and solvent/surface chemistries do affect the onset stress $\sigma_{\mathrm{on}}$ for shear thickening (see Table~\ref{tab:Table_1}), plotting the reduced viscosity $\eta_r$, i.e., $\eta$ normalized by the  viscosity of the suspending  liquid, as a function of  $\sigma$ scaled by $\sigma_{\mathrm{on}}$  reveals remarkably similar  behavior at each volume fraction, with only a few minor deviations.\footnote[1]{There are some discrepancies in the volume fractions, but these are within the typically reported experimental errors associated with determining exact particle density and/or packing fraction~\cite{Poon_2012}. The deviation at low volume fraction for Ref.~\cite{james2018interparticle} is because this data was taken at a constant ratio of [urea]:[solvent] (6 M), meaning that the amount of urea relative to the concentration of surface -CO$_2$H groups changes with packing fraction. The higher relative urea concentration at low packing fractions leads to a lower effective friction coefficient between particle surfaces~\cite{James_2019}.}
Furthermore, these volume-fraction-dependent curves show near quantitative agreement (baseline viscosity values, slope, and extent of shear thickening) with simulations when sets of sliding and rolling friction coefficients $\{\mu_s,\mu_r\}$ were picked that primarily emphasize constraints on sliding, such  as
$\{1,0\}$ or $\{0.5, 0.07\}$.
We note here, and discuss below, that both combinations produce essentially the same curves of $\eta_r(\sigma)$, will lead to the onset of DST at the same packing fraction of $\approx 0.56$, and have the same $\phi_J^\mu \approx 0.58$.
The seeming discrepancy between simulations and experiments at the largest volume fraction shown in Fig.~\ref{fig:Repulsive_Frictional} has its origin in the fact that with experiments it is difficult to map out the multi-valued region in S-shaped $\sigma(\dot\gamma)$ flow curves (Fig.~\ref{fig:Repulsive_Frictional}, inset) that appears for volume fractions beyond the onset of DST.
In the corresponding plots of $\eta(\sigma)$ the bending backward of S-shaped flow curves produces intervals along $\sigma$ where the slope $d$(log$\eta$)/$d$(log$\sigma)$ is larger than 1.
This is reproduced by the simulations, but cannot be observed in conventional rheology experiments, where the slope maximally reaches 1~\cite{Hermes_2016,Han_2019,Saint_2018,Peters_2016,Han_2019, Rathee_2017}.

All of these suspensions are therefore in a regime where their thickening behavior is  dictated primarily by sliding constraints.
We define these suspensions and others that can be collapsed onto similar sets of $\eta_r(\sigma)$ curves as ``standard'' particle suspensions.
The key point here is that while nanometer-scale features and molecular level details at contacting particle surfaces certainly control the interparticle potential (and thus the onset stress), only those that modify sliding constraints appear to affect shear thickening.
The collapse of these experimental data and the agreement with the simulations indicates that at each packing fraction
\begin{enumerate}
\item the stress-dependent balance of frictional versus lubricated contacts is similar,
\item once particles enter into frictional contact, they experience similar microscopic constraints, and
\item these systems primarily experience constraints on sliding  with little penalty for rolling.
\end{enumerate}

\begin{table*}
\centering
\caption{Details of suspension properties in Fig.~\ref{fig:Repulsive_Frictional}}
\begin{tabular}{|c| c| c| c|c|} 
\hline
Particle Identity & Particle size (diameter) & Solvent & Stabilization type & $\sigma_{\mathrm{on}}$ (Pa) \\
\hline
1. PMMA (D'Haene \emph{et al.}) & $0.69~\mu$m & dioctyl phthalate & steric & 220 \\
2. Silica (Royer \emph{et al.}) & $1.54~\mu$m & aqueous glycerol & charge & 418 \\
3. Silica (Egres) & $0.45~\mu$m & PEG-200 & charge & 433\\
4. Carboxylic acid coated Latex (Laun) & $0.20~\mu$m & pH = 6.2 aqueous solution & charge & 5.66\\
5. Carboxylic acid coated PMMA (James \emph{et al.}) & $0.80~\mu$m & 6 M urea in aqueous glycerol & charge & 160\\
6. Glass (Xu \emph{et al.}) & $20~\mu$m & low molecular weight PDMS & charge & 1.66\\
\hline
\end{tabular}
\label{tab:Table_1}
\end{table*}

\subsection{Altering Jamming via Microscopic Frictional Constraints}
\label{sec:jam}
\begin{figure}[!htbp]
\centering
\includegraphics[width=1.0\columnwidth]{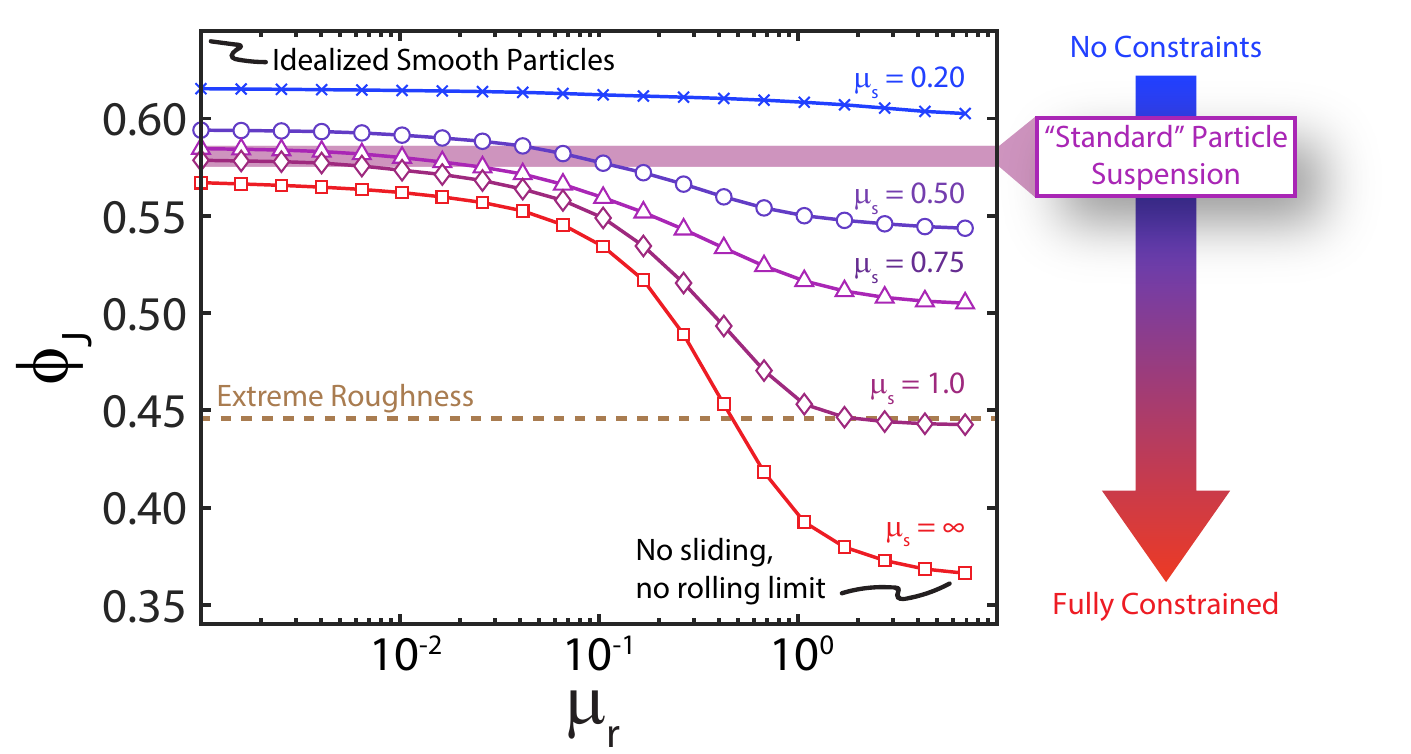}
\caption{
Tuning the jamming volume fraction map by changing stress-activated frictional constraints due to sliding ($\mu_s$) and rolling ($\mu_r$).
The horizontal thick line represents the jamming volume fraction for ``Standard'' Particle Suspensions, while ``Extreme Roughness'' refers to the experimentally measured jamming point for rough raspberry-type particles~\cite{Hsu_2018}.
}
\label{fig:jamming_map}
\end{figure}

As shown in Ref.~\cite{Singh_2020}, the minimum volume fraction $\phi_J^\mu$ required for the onset of jamming depends on both sliding and rolling frictional constraints.
%
%
In  Fig.~\ref{fig:jamming_map} we show this dependence with data traces at various fixed $\mu_s$ while varying $\mu_r$.
The set of friction coefficients $\{\mu_s, \mu_r\}=\{0, 0\}$ represents the limit of no microscopic frictional constraints with a jamming point $\phi_J^\mu \approx 0.65$. Conversely, $\{\mu_s, \mu_r\}=\{\infty, \infty\}$ is the limit where relative particle motion at contact is fully constrained (e.g. contacting particles can neither slide nor roll) and $\phi_J^\mu \approx 0.37$.

The data in Fig.~\ref{fig:jamming_map} provide the context for understanding ``standard'' particle suspensions as well as deviations from it.
The thick horizontal band represents the jamming volume fraction $\phi_J^\mu \approx 0.58$ for ``standard'' particle suspensions shown in Fig.~\ref{fig:Repulsive_Frictional}.
The jamming onset is relatively flat in this region, meaning that friction coefficients for the ``standard'' particle suspensions could range between $0.5<\mu_s<1$ and $0<\mu_r<0.07$, supporting our earlier assertion that in this region shear thickening is primarily due to sliding constraints.
Once $\mu_s \ge 0.5$ and thus significant sliding constraints are present, even small additional amounts of rolling friction lead to large reductions in $\phi_J^\mu$.
This explains observations such as $\phi_J^\mu \approx 0.45$ for raspberry-type particles~\cite{Hsu_2018} or related findings for other rough particles~\cite{Lootens_2005,Hsiao_2017,Pradeep_2020}.
Thus, by experimentally determining $\phi_J^\mu$ one can check immediately whether a shear thickening suspension is governed  mainly by sliding constraints (``standard'') or by a combination of sliding and rolling constraints. 
%
%

\subsection{Deviations from ``Standard'' Behavior due to Rolling Constraints}\label{sec:adh}
We now focus on three illustrative deviations from the ``standard'' suspension behavior, starting with the most intuitive case, namely very rough particles, where large protruding asperities can interlock as particles come into contact.
This reduces the jamming volume fraction significantly~\cite{Lootens_2005,Hsiao_2017, Hsu_2018,Pradeep_2020}, and thus in light of Fig.~\ref{fig:jamming_map} indicates the presence of an additional constraint on rolling.

\begin{figure*}[!htbp]
\centering
\includegraphics[width=1\textwidth]{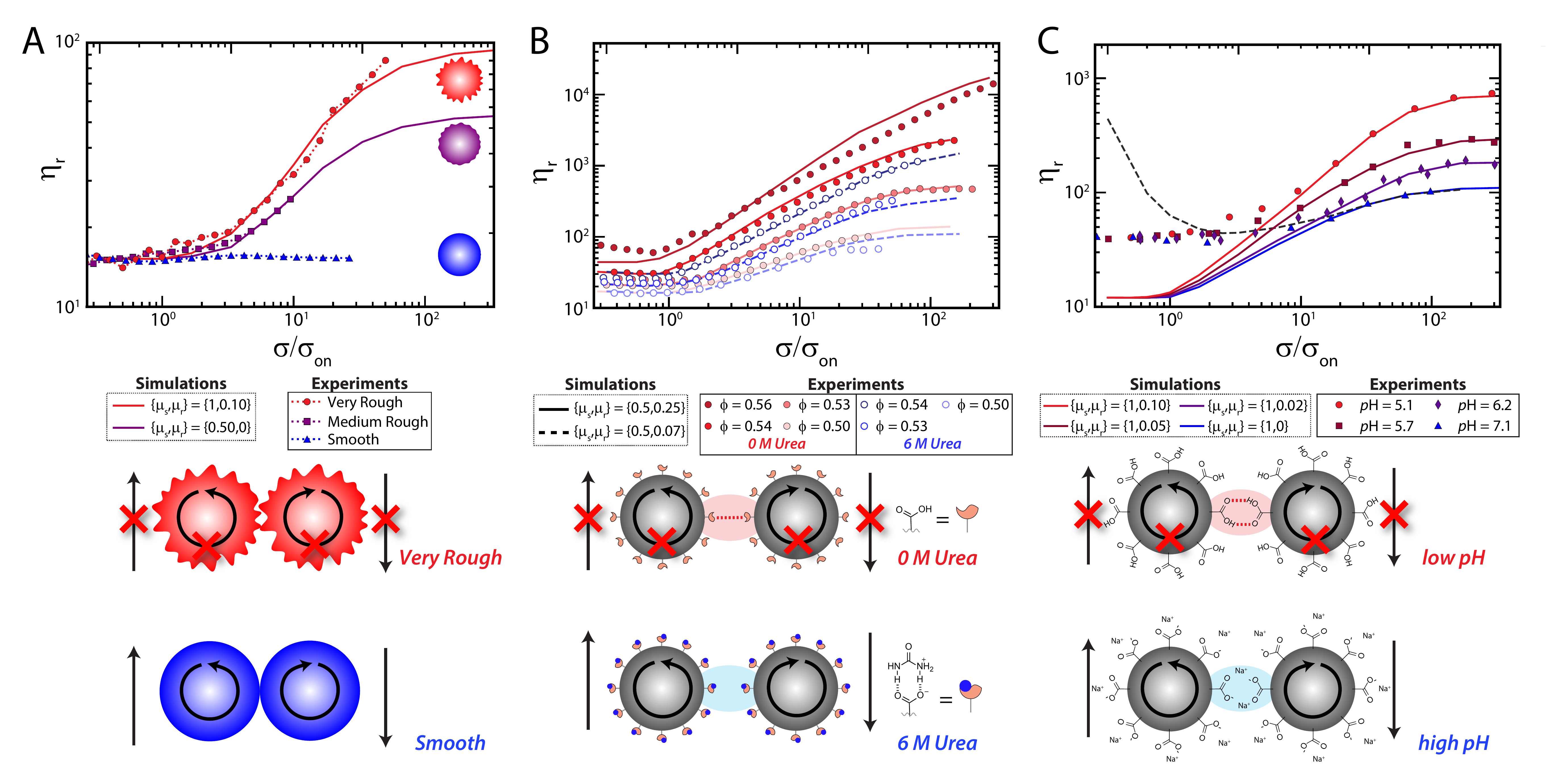}
\caption{
Linking changes in microscopic constraints to deviations from ``standard'' behavior.
(A) Particle surface roughness modifies both sliding and rolling constraints. Experimental data from Ref.~\cite{Hsiao_2017} at volume fraction  $\phi=0.5$ (\emph{symbols}) for variable roughness with $\sigma_{\mathrm{on}}=$ 0.95 Pa (MR) and 5 Pa (VR), simulation data (\emph{lines}) for combinations for $\{\mu_s,\mu_r\}$.
(B) Addition of urea to carboxylic acid (-CO$_2$H) coated particles disrupts interparticle hydrogen bonding and specifically reduces rolling constraints. Experimental data from Ref.~\cite{james2018interparticle}. The onset stress $\sigma_{\mathrm{on}}$ for 0 M and 6M urea concentrations is 5 Pa and 160 Pa, respectively.
(C) Decreasing solution pH for -CO$_2$H coated particles dramatically increases microscopic constraints. Experimental data from Ref.~\cite{Laun_1984} with dashed line being simulation data set from Ref.~\cite{Singh_2019}. Here, $\sigma_{\mathrm{on}}$ is 3500 (pH=7.1), 1925 (6.2), 696.6 (5.7), and 96.3 (5.1) Pa.
}
\label{fig:rolling_friction}
\end{figure*}

As an example we show in Fig.~\ref{fig:rolling_friction}A data from experiments by Hsiao \emph{et al.} with PMMA particles of increasing roughness, from smooth (SM) and medium rough (MR) to very rough (VR) (with dimensionless asperity sizes of 0.026, 0.075, and 0.082, respectively) at a volume fraction $\phi=0.5$ ~\cite{Hsiao_2017}.
For smooth colloids the distance to jamming $\phi_J^{\mu}-\phi$ is so large that no shear thickening is expected and $\eta_r(\sigma)$ is essentially flat. 
For the MR particles the jamming volume fraction was found in the experiments to decrease to $\phi_J^{\mu}=0.59$.
According to  Fig.~\ref{fig:jamming_map} this implies standard suspension behavior, where combinations of $\{\mu_s,\mu_r\}$ for the limit of dominant sliding and negligible friction are appropriate.
Specifically, we find that simulations with $\{\mu_s,\mu_r\}=\{0.5,0\}$  can reproduce the experimental well.
However, for the VR colloids the experimentally determined $\phi_J^{\mu}=0.54$ cannot be reached simply by increasing sliding friction,  even in the limit of infinite $\mu_s$.
This is a clear indicator that now the particle surface is sufficiently rough that asperities interlock to constrain also rolling. 
Based on Fig.~\ref{fig:jamming_map} we expect that $\eta_r(\sigma)$ for the VR suspensions can be reproduced by increasing $\{\mu_s,\mu_r\}=\{0.5,0\}$ used for the MR system to combinations such as $\{1,0.1\}$. 
This is shown by the red trace in  Fig.~\ref{fig:rolling_friction}A.

Changes in particle surface chemistry have also been shown to affect shear thickening.
For example, addition of urea decreases shear thickening in aqueous  suspensions of particles coated with carboxylic acid groups (-CO$_2$H)~\cite{james2018interparticle,James_2019}, as shown in Fig.~\ref{fig:rolling_friction}B.
%
%
In this figure the experimental data have been scaled by their respective onset stresses for thickening $\sigma_{\mathrm{on}}$.
%
Data without added urea (i.e. 0 M) are matched by the simulations only when combinations of sliding and rolling friction are used that fall outside the ``standard'' range in Fig.~\ref{fig:jamming_map}, such as $\{\mu_s,\mu_r\}=\{0.5,0.25\}$ or $\{0.6,0.10\}$.
Such large effective friction, resulting from significant constraints on sliding plus additional rolling resistance, can be be attributed to the formation of hydrogen bonds between carboxylated surfaces as particles are coming into close contact.
The reduced shear thickening in 6 M urea is then caused by ``capping'' of the -CO$_2$H groups.
This disrupts hydrogen bonding, and urea thereby reduces the effective friction coefficient to levels of ``standard'' suspensions.
This is seen by the fact that we can model the 6 M urea data in Fig.~\ref{fig:rolling_friction}B with combinations such as $\{\mu_s,\mu_r\}=\{0.5,0.07\}$.
%


The constraint-based perspective also sheds new light on the observation that shear thickening for CO$_2$H-coated latex is highly dependent on pH~\cite{Laun_1984}.
At a constant volume fraction, suspensions at pH = 7.1 exhibit continuous shear thickening whereas those at pH = 5.1 display DST (Fig.~\ref{fig:rolling_friction}C).
While the viscosity of the low stress state of the experimental data is considerably higher than that of the simulation, this is due to the existence of a finite yield stress and shear thinning in the experimental data set (here we only show the data for intermediate to high stress region).
We can easily account for that in the simulations by including attractive forces between particles, similar to what is depicted in Fig.~\ref{fig:Sketch} and discussed in detail in Ref.~\cite{Singh_2019}.
By fixing $\mu_s=1$ for convenience and increasing rolling constraints $\mu_r$ from 0 to 1.0 in the simulations, we can quantitatively reproduce the experimental data.
The constraint-based picture therefore allows us to move beyond the results of Noy \emph{et al.}~\cite{Noy_1997}, who showed that the effective friction coefficient between -CO$_2$H coated surfaces increases with decreasing pH.
Specifically, increased hydrogen bonding between ``sticky'' or adhesive protonated -CO$_2$H groups at lower pH gives rise to increased sliding and rolling constraints,  which are responsible for the changes in the shear thickening behavior observed on macroscopic scales.

\section*{Conclusions}
%
%
%

Our numerical simulations highlight the power of a constraint-based approach for understanding and predicting the shear thickening behavior of diverse kinds of dense suspensions, including a wide range of different particle types and particle surface features.
We find that two constraints on relative particle movement  suffice for quantitative modeling, namely separate constraints on sliding and rolling that are activated when the local stress exceeds a threshold such that particles are coming into direct contact and experience friction forces.
A central point in this approach is that each constraint can represent a variety of different physical or chemical interactions giving rise to this friction.
In particular, the approach allows one to translate ideas about chemical interactions typically developed under more dilute, closer to equilibrium conditions to these concentrated, out-of-equilibrium systems.
In the simulations, constraints are implemented via  associated friction coefficients $\mu_s$ and $\mu_r$, whose magnitudes provide an indicator of their relative strength.
The combination of sliding and rolling constraints can affect the shear thickening behavior in highly nonlinear ways, where already small amounts of additional rolling friction can make an outsized contribution.
As Fig.~\ref{fig:jamming_map} demonstrates, for physically realistic values $\mu_s \approx 0.5$ additional rolling resistance can have a very large effect on $\phi_J^\mu$ and thus on the stress response of a dense suspension.
As a result, plots of the suspension viscosity $\eta(\sigma)$ as a function of applied shear stress can be viewed as a macroscopic reporter of stress-activated constraints that originate from contact interactions at the nanoscale.
In particular:
\begin{itemize}
\item if $\eta(\sigma)$ \emph{follows} the ``standard'' rheology in Fig.~\ref{fig:Repulsive_Frictional}, then the rolling and sliding friction coefficients fall within the ``flat'' region of $\phi_J^\mu$ in Fig.~\ref{fig:jamming_map}, indicating minimal surface roughness and a lack of strong interparticle adhesive interactions for the given combination of particle/solvent chemistry;
\item if $\eta(\sigma)$ \emph{deviates} from the ``standard'' rheology, then the rolling and/or sliding constraints are stronger due to surface roughness or adhesive interactions such as hydrogen bonding between particle surfaces 
\end{itemize}

We note that access to such detailed information about the relative contributions from sliding and rolling is exceedingly difficult to obtain from experiments that use scanning probe techniques, which necessarily focus on lateral sliding motion.

In terms of an effective coefficient of friction accounting for the role of both $\mu_s$ and $\mu_r$, the ``flat'' region of $\phi_J^\mu$ in Fig.~\ref{fig:Repulsive_Frictional}  provides an explanation as to why prior simulations that did not include rolling friction were still able to model ``standard'' suspensions as long as they used large values of $\mu_s \approx 1$~\cite{Mari2015discontinuous,Ness_2016, Singh_2018}. 
More generally, Fig.~\ref{fig:Repulsive_Frictional} shows that  combinations of different types and strengths of contact interactions, represented by different sets of values for $\mu_s$ and $\mu_r$, can give rise to the same jamming threshold $\phi_J^\mu$ and thus the same shear thickening behavior.
Therefore, our results suggest that $\phi_J^\mu$ suitably represents the combined effect of the two different types of constraint. 
%
The same framework could also be adapted to understand how stress-deactivated constraints~\cite{Guy_2018} give rise to shear thinning.
Finally, while we here only focused on the connection between microscopic constraints and the macroscopic shear thickening response, there are open questions on how different stress-activated constraints alter the mesoscale force chain network.
%
%
%
%
%
%

\begin{acknowledgments}
We acknowledge support from the Center for Hierarchical Materials Design (CHiMaD) under award number 70NANB19H005 (US Dept. Commerce). 
and from the Army Research Office under grants W911NF-19-1-0245, W911NF-20-2-0044, and W911NF-21-1-0038. 
\end{acknowledgments}


\bibliography{dst}

\clearpage
\begin{widetext}
\begin{appendix}

\section*{\large Supplemental Material for ``Stress-activated Constraints in Dense Suspension Rheology"}
In this document we provide details about the simulation scheme used in the simulations. 

\section*{Methodology}\label{rol_fric}
For all of the experimental systems studied both the Reynold's Number and the Stokes number are vanishingly small as they both scale with the square of the particle size.  
This assumption allows us to simulate particle suspensions in the overdamped limit.
We simulate an assembly of inertialess frictional spheres in a Newtonian viscous fluid under an imposed stress $\sigma$, that gives rise to an imposed velocity field $\vec{v} = \dot\gamma(t)\hat{\vec{v}}(\bm{x}) = \dot\gamma(t)(x_2, 0, 0)$.
We use Lees-Edwards periodic boundary conditions with $N=2000$ particles in a unit cell with a bidisperse particles, with radii $a$ and $1.4a$ mixed at equal volume fractions to avoid ordering~\cite{Mari_2014}.
The particles interact through short-range hydrodynamic forces (lubrication), short-ranged repulsive forces (electrostatics), and frictional contact forces.
The equation of motion for $N$ spheres can be reduced to $6N$-dimensional force (as well as torque) balance between the 
hydrodynamic ($\vec{F}_{\mathrm{H}}$), repulsive ($\vec{F}_{\mathrm{R}}$), and contact ($\vec{F}_{\mathrm{C}}$) interactions according to:
\begin{equation}
  \vec{0} = \vec{F}_{\mathrm{H}}(\vec{X},\vec{U}) + \vec{F}_{\mathrm{C}}(\vec{X}) + \vec{F}_{\mathrm{R}}(\vec{X}), 
  \label{eq:force_balance}
\end{equation}
where $\vec{X}$ and $\vec{U}$ denote the particle positions  and velocities/angular velocities, respectively.
To match simulations to the experimental data we tuned the Debye length $\lambda$ in the repulsive force $F_R$ and the friction coefficients $\mu_r$ and $\mu_s$ in $F_C$.  
For more information on the simulation method, including the functional forms of the forces in the equations of motion as well how they are solved numerically, we refer the reader to refs. ~\cite{Mari_2014, Singh_2018,Singh_2020}. 
Unit scales
are $\dot{\gamma}_0 \equiv F_{\rm 0}/{6\pi \eta_0 a^2}$ 
for the strain rate and $\sigma_0 \equiv \eta_0 \dot{\gamma}_0 = F_{\rm 0}/6\pi a^2$ for the stress.
The onset stress for shear thickening, $\sigma_\mathrm{on}$, used in the main text is obtained from experimental data and scales as $\sigma_\mathrm{on} = \alpha \sigma_0$, with typical values for $\alpha$ around unity.

\end{appendix}
\clearpage
\end{widetext}

\end{document}